\documentclass[fleqn,usenatbib]{mnras}

\usepackage{newtxtext,newtxmath}

\usepackage[T1]{fontenc}

\DeclareRobustCommand{\VAN}[3]{#2}
\let\VANthebibliography\thebibliography
\def\thebibliography{\DeclareRobustCommand{\VAN}[3]{##3}\VANthebibliography}

\usepackage{tikz,xcolor,hyperref}

\usepackage{graphicx}	
\usepackage{amsmath}	
\usepackage{gensymb}
\usepackage{longtable}
\usepackage{anyfontsize}
\usepackage{siunitx}

\newcommand{\hb}{H$\beta$}

\newcommand{\ratio}{$L_{\mathrm{H}\beta}/L_{[\mathrm{OIII}]}$}

\newcommand{\refbf}{}
\newcommand{\refbftwo}{}

\defcitealias{burke_characteristic_2021}{B21}

\definecolor{lime}{HTML}{A6CE39}
\DeclareRobustCommand{\orcidicon}{%
    \begin{tikzpicture}
    \draw[lime, fill=lime] (0,0) 
    circle [radius=0.16] 
    node[white] {{\fontfamily{qag}\selectfont \tiny ID}};
    \draw[white, fill=white] (-0.0625,0.095) 
    circle [radius=0.007];
    \end{tikzpicture}
    \hspace{-2mm}
}

\newcommand{\orcidChrisO}{\href{https://orcid.org/0000-0003-0017-349X}{\orcidicon}}
\newcommand{\orcidChrisW}{\href{https://orcid.org/0000-0002-4569-016X}{\orcidicon}}
\newcommand{\orcidJT}{\href{https://orcid.org/0000-0003-2858-9657}{\orcidicon}}
\newcommand{\orcidNA}{\href{https://orcid.org/0009-0005-7553-049X}{\orcidicon}}
\newcommand{\orcidRW}{\href{https://orcid.org/0000-0002-5325-2709}{\orcidicon}}

\newcommand{\orcidAT}{\href{https://orcid.org/0009-0004-8522-152X}{\orcidicon}}

\title[Variability of low-luminosity AGN]{Optical Variability Structure Function of Low-Luminosity AGN using ATLAS Lightcurves}

\author[A. Tan et al.]{
Ashley Hai Tung Tan,$^{1}$\thanks{E-mail: ashleyhaitung.tan@anu.edu.au}\orcidAT
Christian Wolf,$^{1,2}$\orcidChrisW
Neelesh Amrutha,$^{1}$\orcidNA
Christopher A. Onken,$^{1,2}$\orcidChrisO
\newauthor{}
John L. Tonry$^{4}$\orcidJT
and Rachel Webster$^{3}$\orcidRW
\\
$^{1}$Research School of Astronomy and Astrophysics (RSAA), Australian National University, Canberra ACT 2611, Australia\\
$^{2}$Centre for Gravitational Astrophysics (CGA), Australian National University, Building 38 Science Road, Acton ACT 2601, Australia \\
$^{3}$School of Physics, University of Melbourne, Parkville, Victoria 3010, Australia \\
$^{4}$Institute for Astronomy, University of Hawaii, 2680 Woodlawn Drive, Honolulu, HI 96822-1897, U.S.A. \\
}

\date{Accepted XXX. Received YYY; in original form ZZZ}

\pubyear{2026}

\begin{document}
\definecolor{asparagus}{rgb}{0.53, 0.66, 0.42}
\label{firstpage}
\pagerange{\pageref{firstpage}--\pageref{lastpage}}
\maketitle

\begin{abstract}
The origin of the optical flux variability in active galactic nuclei (AGN) is largely unknown. Previous studies have correlated features of the variability structure function (SF) with AGN properties, though they mostly involved high-luminosity AGN to avoid biases from host galaxy flux. In this work, we characterise optical variability in a sample of 246 low-luminosity AGN at $z < 0.1$ from the Six-degree Field Galaxy Survey (6dFGS) through the ensemble variability SF. We use lightcurves from the Asteroid Terrestrial-impact Last Alert System (ATLAS) with a cadence of $\sim$2 days over eight years, and perform host-AGN decomposition on recent spectra to obtain the host fraction. We find that the slope of the SF depends on black hole mass, increasing from $\sim 0.1$ at $\log M_{\mathrm{BH}}/M_\odot \sim 6.5$ to $\sim 0.3$ at $\log M_{\mathrm{BH}}/M_\odot \sim 8$. Contrary to some earlier work, \refbftwo{we} do not find breaks in the SF, and two-epoch spectra taken $\sim$20 years apart suggest that the SF keeps rising into decadal timescales. In addition, we measure an anticorrelation of the amplitude with the luminosity and a positive correlation with the black hole mass. The variability behaviour also suggests that extinction is not the main driver of the variety in Seyfert subtypes. 
\end{abstract}




\section{Introduction}
Active Galactic Nuclei (AGN) are known to have non-periodic variability across all wavelengths. This characteristic variability, intrinsic to the AGN, has enabled the detection of AGN and is widely used to probe their inner structure and processes. A significant component of this variability is thought to originate from accretion disc instabilities, for example through magneto-rotational instabilities (MRI) and convective instabilities \citep{balbus_powerful_1991,jiang_opacity-driven_2020}. The reprocessing of X-ray variability is likely not the dominant mechanism behind long-term optical variability, as growing evidence points to a lack of correlation between X-ray and UV lightcurves \citep{arevalo_correlation_2009,edelson_first_2019,kara_xrayreverb_2023} and X-ray variability amplitudes that are too small to account for the observed UV/optical amplitudes \citep{hagen_modelling_2023,beard_testing_2025}. If the variability processes were well understood, there might be great potential in using flux variability, easily available from time-domain surveys such as the upcoming Legacy Survey in Space and Time \citep[LSST;][]{abell2009lsst,ivezic_lsst_2019}, to serve as a complementary probe to spectroscopy for AGN properties. Since the physical origin of the optical variability remains uncertain, many studies have explored correlations between variability behaviour and AGN parameters to constrain the underlying mechanisms.

Optical variability is either characterised with structure functions  \citep[SF;][]{collier_characteristic_2001,de_vries_structure_2005,macleod_modeling_2010} or the power spectrum density \citep[PSD;][]{kelly_stochastic_2011,kelly_active_2013,simm_pan-starrs1_2016,arevalo_universal_2024}. Both probe the variability amplitude as a function of the time interval between measurements, but the SF is thought to be better suited for unevenly sampled time series data \citep{collier_characteristic_2001}. An early study by \citet{kelly_are_2009} found that their dataset consisting of 100 optical light curves is well fit by a damped random-walk (DRW) model. Subsequently, the DRW model has been used by many studies to parametrise their data. In this model, the SF has a slope of $\gamma=0.5$ up to a break at the decorrelation timescale $\tau_{\mathrm{break}}$. Above $\tau_{\mathrm{break}}$, the data points are decorrelated and $\gamma=0$. While many studies have found a decent agreement with the DRW model \citep{macleod_modeling_2010,macleod_description_2012,kozlowski_revisiting_2016,tang_universality_2023,tang_vsf_2024}, some have found slopes that deviate from it \citep{vanden_berk_ensemble_2004,caplar_optical_2017,sanchez-saez_quest-silla_2018,wang_deviation_2019,arevalo_universal_2024}. Some works argue that the shape of the SF or PSD is not well-described by a simple power law, such as \citet{arevalo_universal_2024}, which found that their power spectra have a bending power law shape. This could indicate that the DRW model does not fully model AGN variability and a more complex model is required, such as a damped harmonic oscillator model, a continuous-time autoregressive moving average model or a higher-order autoregressive moving average process \citep{kelly_flexible_2014,simm_pan-starrs1_2016,yu_examining_2022,yu_examiningii_2025}. 

Besides the slope, studies often measure the correlation of the variability amplitude with AGN properties such as luminosity, black hole (BH) mass, accretion rate/Eddington ratio, and the rest-frame wavelength, which indicates the region of the accretion disc probed. It is generally agreed that the variability amplitude is anticorrelated with the AGN luminosity \citep{kelly_are_2009,macleod_modeling_2010,kozlowski_revisiting_2016,caplar_optical_2017,de_cicco_structure_2022,tang_universality_2023}, though studies disagree on whether there is a dependence on black hole mass $M_{\mathrm{BH}}$, such as a negative correlation with $M_{\mathrm{BH}}$ by \citet{kelly_are_2009}, no correlation by \citet{de_cicco_structure_2022}, and a positive correlation by \citet{kozlowski_revisiting_2016}. Note, that a slope dependence on black hole mass implies that any amplitude-mass dependence depends on the probed timescale. These correlations were used to constrain the mechanism driving the variability; for example \citet{yu_examining_2022} related the correlation with accretion rate to the size of the X-ray corona, while \citet{son_temperature_2025} attempted to correlate the variability and rest-frame wavelength to constrain the disc temperature profile. 

Studies have also interpreted the break timescales as characteristic variability timescales, though many have found conflicting relations between the break timescale and AGN properties. \citet{kelly_are_2009} found that their characteristic timescales depend on luminosity and mass in a manner that is consistent with the disc dynamical or thermal timescales, and suggested that the variability is driven by thermal fluctuations in the accretion disc. They proposed that at short timescales, the disc cannot fully react to random fluctuations, resulting in suppressed variability. \citet[][B21 hereafter]{burke_characteristic_2021} found that the break timescale only depends on black hole mass with $\tau_{\mathrm{break}} \propto M_{\mathrm{BH}}^{0.38\pm0.05}$, and proposed this as a method to measure masses. Other works have also found varied correlations, such as $\tau_{\mathrm{break}} \propto M_{\mathrm{BH}}^{0.65-0.55}R_{\mathrm{Edd}}^{0.35-0.3}$ \citep{arevalo_universal_2024}, and $\tau_{\mathrm{break}} \propto M_{\mathrm{BH}}^{0.6-0.8}$ \citep{su_bhmass_2024}. A limited sample size and sampled parameter space could have made it difficult for studies to remove the influence of other variables and properly measure a dependence with just the black hole mass. Apart from that, the black hole mass is also a parameter with high uncertainty. The method used to estimate most black hole masses in the literature, the single-epoch virial mass method, relies on the radius-luminosity (R-L) relation to obtain the broad line region (BLR) radius from a single spectrum. However, this method has an uncertainty of $\sim0.5$ dex \citep{shen_mass_2013} due to the uncertainty in the virial factor and object-to-object variations. In addition, recent studies have found that mass estimates are biased by $R_{\mathrm{Edd}}$ \citep{du_rm_2014,gravity_collaboration_size-luminosity_2024} and \ratio  \citep{amrutha_masscorrection_2026}, further exacerbating this issue. Another factor that could contribute to disagreeing results is the use of sparsely sampled data with short baselines relative to the measured characteristic timescales, especially in earlier studies. For example, \citet{stone_optical_2022} found that even with baselines up to 10-15 years, the measured break timescales increase with baseline length, suggesting that shorter baselines in the literature are insufficient to probe the underlying timescales. However, the emergence of large time-domain surveys in recent years enabled studies to tap into datasets with excellent cadence and longer baselines. 

Using these survey datasets, a few studies have found that scaling the SF or PSD with the orbital or thermal timescales resulted in a universal SF or PSD that well-characterises the sample over a wide range in their properties. \citet{tang_universality_2023,tang_vsf_2024} used 5\,000 lightcurves from the \refbftwo{NASA} Asteroid Terrestrial-impact Last-Alert System \citep[ATLAS;][]{tonry_atlas_2018} with a cadence of 1-2 days spanning more than five years to construct ensemble SFs of the brightest quasars. When they scaled the time interval by the thermal timescale of the accretion disc, they found a universal structure function independent of AGN properties in the form of a single power law ($\gamma=0.5$) with no breaks. However, it is worth noting that they excluded data redder than rest-frame 3000\AA \ where the amplitude dependence on the rest-frame wavelength deviated from a simple relation. Interestingly,  \citet{caplar_optical_2017} have also previously found similar correlations of the amplitude with rest-frame wavelength. Another work, \citet{arevalo_universal_2024}, used Zwicky Transient Facility \citep[ZTF;][]{bellm_ztf_2019} lightcurves to probe $\sim 2900$\AA \ rest-frame. Contrary to \citet{tang_universality_2023}, they found PSDs well-parametrised by a bending power law. They found that scaling the frequency of their power spectra using $M^{\mathrm{0.65-0.55}}R_{\mathrm{Edd}}^{0.35-0.3}$, which corresponds to the scaling of the orbital timescale at $R_g$, yields universal power spectra with the same bend frequency. While these results suggest that the variability processes are related to the orbital or thermal timescales, the mechanism behind these observations is still not well understood. 

It is important to note that most of the literature has studied quasars, where emission from the host galaxy is negligible and there is no need to correct the measured SF to obtain the intrinsic AGN SF. Only a handful of Seyferts have been the focus of variability studies, such as by \citet{kasliwal_are_2015}. So, it could be the case that the correlations found in the literature only apply to high-luminosity AGN. There are differences in the structure and processes in the disc and corona in different luminosity regimes, as evidenced by the fraction of luminosity emitted in X-rays depending on the overall luminosity as expressed by the $L_\mathrm{X}-L_{\mathrm{UV}}$ relation \citep{vignali_lxluv_2003}. \citet{liu_observational_2021} also found a dependence of the optical-to-X-ray power-law slope parameter $\alpha_{\mathrm{OX}}$ with black hole mass and Eddington ratio. Since the disc–corona coupling influences both the disc emission and any reprocessed X-ray variability, the variability mechanisms in low-luminosity AGN may therefore differ from those in their high-luminosity counterparts. Optical photometry of high-redshift, high-luminosity objects also probes the rest-frame UV corresponding to the inner, hotter disc, which may not necessarily have the same variability processes as the outer, cooler disc emitting in the rest-frame optical. Some studies even suggest a breakdown of the accretion disc structure at extremely low Eddington ratio \citep{hagen_collapse_2024,kang_collapse_2025} or a two-zone accretion disc with a thin and slim disc component \citep{li_twozone_2024}, which may affect disc timescales and variability behaviour. Studies using Kepler light curves \citep{mushotzky_kepler_2011,kasliwal_are_2015,smith_kepler_2018} probed low-luminosity AGN, but they did not apply a host correction and Kepler is known to suffer from instrumental issues \citep{caldwell_kepler_2016,van_cleve_kepler_2016,moreno_properties_2021}. Therefore, studying the optical variability of low-luminosity AGN will reveal the behaviour in a mostly unexplored region of parameter space. If the high-luminosity correlations extend to the low-luminosity end, we expect to be able to measure relevant characteristic timescales with surveys such as ZTF, ATLAS, and \refbftwo{soon} LSST. For example, the break timescale predicted by \citetalias{burke_characteristic_2021} for our sample at the median mass of $\log M_{\mathrm{BH}}/M_\odot = 7.0$ is $\sim50$ days, which is easily measurable with our eight-year long lightcurves if present. 

This work characterises the optical variability of a sample of low-luminosity, broad-line AGN at $z<0.1$ using the ensemble variability structure function. We use a sample of low-luminosity AGN derived from a catalogue of broad-line AGN \citep{hon_belagn_25} from the Six-degree Field Galaxy Survey \citep[6dFGS;][]{jones_saunders_2004_6dfgs,jones_read_2009_6dfgs} and high-cadence light curves from \refbftwo{NASA ATLAS}.
To measure AGN properties, we perform host-galaxy decomposition on spectra taken using the Wide-Field Spectrograph \citep[WiFeS;][]{dopita_hart_2007} on the ANU 2.3m telescope \citep{mathewson_2.3m_2013}. Throughout this paper, we use AB magnitudes and adopt a flat Lambda cold dark matter cosmology with $\Omega_\Lambda = 0.7$ and $H_0=70 \ \mathrm{km \ s^{-1} \ Mpc^{-1}}$.

\section{Sample and spectra}\label{sec:sample_lc_data}
\begin{figure*}
    \centering
    \includegraphics[width=\linewidth]{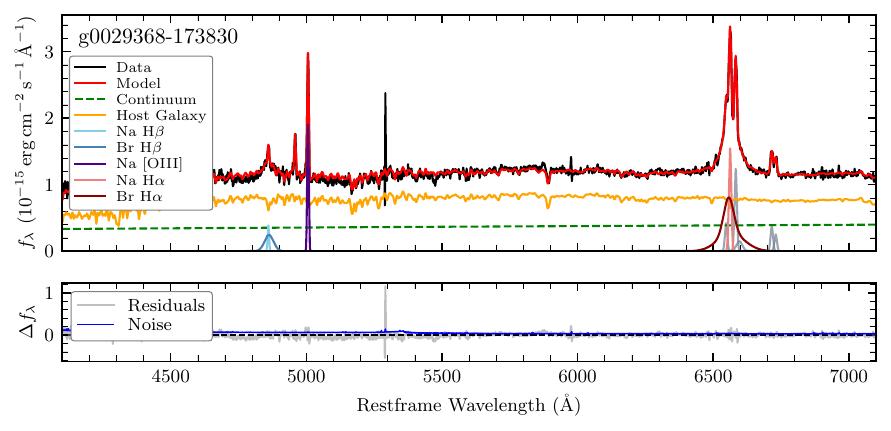}
    \caption{A spectrum decomposition of g0029368-173830 using BADASS3. The top panel shows the spectrum with the fitted components as detailed in the legend. Additional lines indicate the narrow [NII] and [SII] lines. The bottom panel shows the residuals and the noise.}
    \label{fig:spec_fit}
\end{figure*}
The sample was selected from the broad-line AGN catalogue by \citet{hon_belagn_25} from 6dFGS, which is complete at a magnitude of $K<12.65$, except for bright quasars as point sources are avoided by 6dFGS. From this parent sample, we limit the sample to objects at $z<0.1$ to select low-luminosity AGN. The difference lightcurves may be contaminated by flux from neighbouring objects, so objects with neighbours within 4" were excluded, along with objects with $E(B-V)>0.3$ to exclude objects heavily affected by galactic extinction. 

To perform host galaxy correction, we use recent spectra taken with WiFeS on the ANU 2.3m telescope. WiFeS is an integral field spectrograph with a $38^{\prime\prime} \times 25^{\prime\prime}$ field-of-view and typically binned to 1$^{\prime\prime}$ spaxels. The spectra were taken using the B3000 and R3000 gratings with the RT560 dichroic, then reduced using PyWiFeS \citep{childress_vogt_2014_pywifes}. They were extracted using a 6.7$^{\prime\prime}$ aperture to be consistent with the 6dFGS fibre size for a separate study comparing the AGN mass estimates in two epochs \citep{amrutha_masscorrection_2026}. 

\begin{figure}
    \centering
    \includegraphics[width=\linewidth]{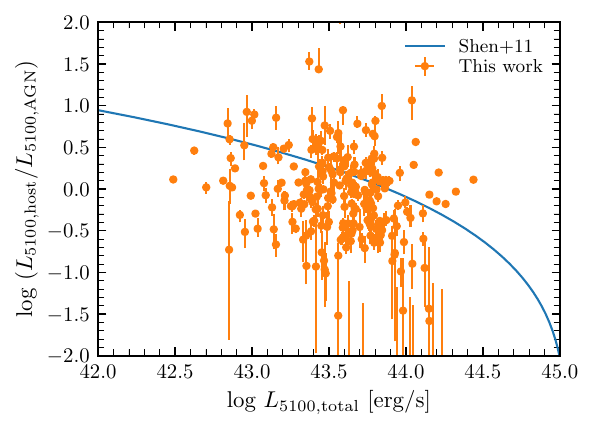}
    \caption{The flux ratio of the host galaxy and AGN at rest-frame 5100\AA \ plotted against the total luminosity. A line indicates the relation by \citet{shen_catalog_2011}. The spectrum decomposition results fit the relation on average. }
    \label{fig:shen_host}
\end{figure}

The spectra were fit using the Bayesian AGN Decomposition Analysis for SDSS Spectra \citep[BADASS3;][]{sexton_bayesian_2021}. BADASS3 is an open source code developed to simultaneously fit all spectral components using a Markov-Chain Monte Carlo approach. All spectra were fit using a rest-frame wavelength range of 4000\AA-6800\AA, and we allow the algorithm to fit an AGN power law, a host template, a Fe II optical continuum template, and two Gaussians for each broad (>1200 km/s) and narrow (<1200 km/s) emission line. Emission lines that were fitted include broad and narrow H$\alpha$ and H$\beta$, as well as narrow [OIII], [SII], and [NII]. The H$\gamma$ line was masked and not fitted. The host templates used in the fitting process are single-stellar population (SSP) templates from the E-MILES library \citep{vazdekis_uv-extended_2016}. To minimise possible degeneracy between the bluer host templates and the AGN continuum, we visually compare turn-off CLAGN spectra from the same parent sample \citep{amrutha_clagn_24}, which no longer show AGN continua, with the host galaxy templates. This comparison indicates that the host stellar populations are generally consistent with ages $\geq 6$ Gyr, which we therefore adopt as a lower age limit in the fitting. A spectral fit to one of the objects in the sample, g0029368-173830, is shown in Figure \ref{fig:spec_fit}. A more detailed description of the spectral fitting process will be described in Amrutha et al. (in prep.). We compare the fitted host fractions with those from the SDSS DR7 quasar catalogue \citep{shen_catalog_2011} in Figure \ref{fig:shen_host}, where $L_{5100}$ is the monochromatic luminosity, $\lambda L_\lambda$ at 5100\AA. Our measured host fractions approximately follow the relation with some scatter. Due to the aforementioned degeneracies in the bluer host templates and AGN continua, we expect some uncertainties in the host estimates. The host correction was performed using the median host in each SF sample bin to reduce the effects of these uncertainties. We apply further cuts to the sample based on the spectral fits, and exclude objects where the broad \hb \ line has a weaker signal than the average noise level. We also remove objects with double-peaked broad emission lines, which are possible indicators of binary supermassive black holes \citep{decarli_bsmbh_2013}. 

The black hole masses were estimated using the broad H$\beta$ line luminosity and FWHM \citep{vestergaard_peterson_2006}. The Eddington ratio for the sample was estimated by converting $L_{5100}$ to $L_{\mathrm{bol}}$ using a bolometric correction factor of 9.26 \citep{richards_spectral_2006}. In Section \ref{subsec:smss_calib}, we use photometry from the SkyMapper Southern Survey, a multi-epoch, multi-colour survey of the entire southern sky \citep{wolf_onken_2018_smss,onken_skymapper_2024}, to calibrate the AGN luminosity and black hole mass. The survey was taken using the SkyMapper telescope, a 1.3m telescope at Siding Spring Observatory near Coonabarabran. Therefore, we also exclude objects that do not have good quality SkyMapper photometry in the $g$ and $r$ band. 

\begin{figure}
    \centering
    \includegraphics[width=0.925\linewidth]{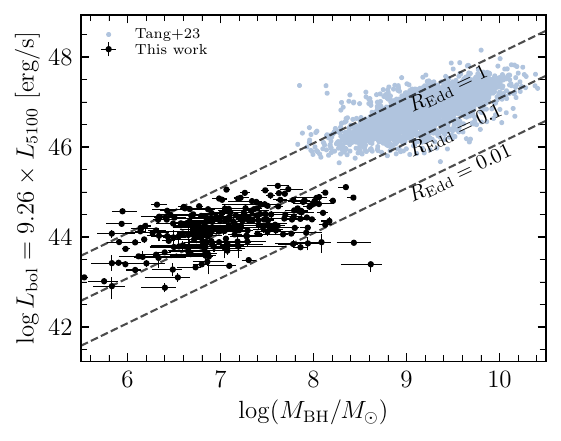}
    \caption{\refbf{Comparison of $L_{5100}$ luminosity and black-hole mass in our sample with that in \citet{tang_universality_2023}. Error bars are derived from spectral fitting, though true uncertainties are larger due to unknown viewing angles, anisotropies, and the single-epoch mass estimation method. Dashed lines correspond to Eddington ratios of 0.01, 0.1, and 1. Note that black hole masses in the two samples were estimated from different emission lines.}}
    \label{fig:mass_lum_compare}
\end{figure}

The final sample, after additional cuts applied to the lightcurves as detailed in Section \ref{sec:lc_data} and \ref{sec:sf_methods}, consists of 246 objects with a median luminosity of $\log L_{5100}/(\mathrm{erg\ s^{-1}}) \sim 43.3$ and median black hole mass of $\log M_{\mathrm{BH}}/M_\odot \sim 7.0$. The median Eddington ratio of the sample is $\sim 0.1$, and the median redshift is 0.06. A comparison of our sample's parameter space with the sample from a previous work probing the brightest quasars \citep{tang_universality_2023} is shown in Figure \ref{fig:mass_lum_compare}. We converted $L_{5100}$ in our sample to $L_{\mathrm{bol}}$ for comparison. Note, that the black hole masses in the high luminosity sample have been estimated using different emission lines due to the redshift range \citep{rakshit_stalin_2020}.  

\section{Lightcurve data}\label{sec:lc_data}
To characterise the variability, we use difference lightcurves from \refbftwo{NASA ATLAS \citep{tonry_atlas_2018}}, which is a system of four telescopes that survey the entire sky with a cadence of 1-2 days to detect dangerous near-Earth asteroids. The two northern ATLAS units are located in Hawaii at Haleakala and Mauna Kea, while the two southern units are located in El Sauce, Chile, and Sutherland, South Africa. The two southern ATLAS units started operations in 2022, so we omit objects at $\delta \leq-50^{\circ}$ to standardise the baseline and avoid varying window effects. 

We also exclude known Changing-Look AGN \citep[CLAGN;][]{lamassa_cales_2015,ricci_changing-look_2023} and CLAGN candidates by identifying lightcurves that abruptly transition between strongly variable and weakly variable. This is because CLAGN often exhibit drastic variability changes when they "turn on" or "turn off", which may be a separate process from the variability we probe. The removed objects include 19 confirmed CLAGN \citep{amrutha_clagn_24} and nine CLAGN candidates identified visually. ATLAS observes in two bands, the cyan band ($\lambda_{\mathrm{pivot}}=529 \ \mathrm{nm}$, $W_\mathrm{eff} = 214 \ \mathrm{nm}$) and the orange band ($\lambda_{\mathrm{pivot}}=675 \ \mathrm{nm}$, $W_\mathrm{eff} = 237 \ \mathrm{nm}$). Observations in the orange band are carried out every one or two days, weather permitting, while the cyan band is only observed in the dark phase of the moon. We primarily utilise orange band data in the subsequent structure function analysis. 

Because a new field corrector lens was installed while the software continued to use the same astrometric solution, affecting image quality before MJD 57900, we use lightcurve data from MJD 57900 to MJD 60815. Changes in the reference image (wallpaper) used to produce difference images also resulted in a discontinuity in some lightcurves at MJD 58882. Thus, we do not sample pairs across that date, resulting in a maximum effective baseline of $\sim$ 1900 days in the rest-frame. An earlier wallpaper update at MJD 58417 did not appear to have produced any discontinuity. 

\begin{figure*}
    \centering
    \includegraphics[width=\linewidth]{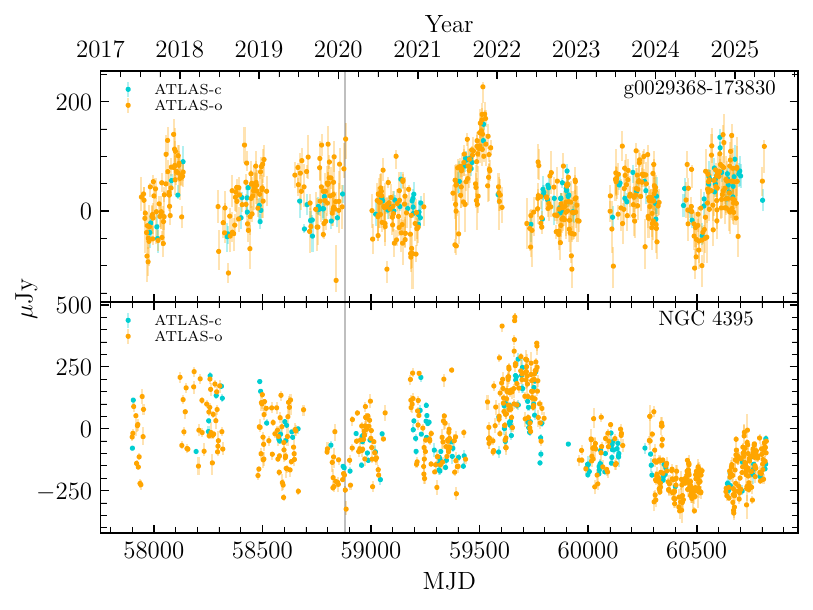}
    \caption{The one-day stacked ATLAS \refbf{difference} lightcurves of g0029368-173830 (top panel) and NGC4395 (bottom panel). \refbftwo{These lightcurves are difference lightcurves based on a fixed "wallpaper" as explained in Section \ref{subsec:smss_calib}, such that both the nuclear host component and the mean brightness of the AGN are subtracted. 
    } The vertical line indicates the date of the wallpaper changes at MJD 58882.}
    \label{fig:example_lc}
\end{figure*}

To increase the signal-to-noise ratio, we create one-day stacks of the orange and cyan data. Each target is scheduled for four exposures per night, so nights with fewer exposures, likely due to poor weather, are excluded from the analysis. We also used an error cutoff of 45$\mu\mathrm{Jy}$ determined from iterative sigma clipping to discard data points with high errors. After cleaning, the lightcurves have an average cadence of $\sim$2 days. A more detailed description of the lightcurve processing is described by \citet{amrutha_clagn_24}. The stacked \refbf{difference} lightcurves of two example objects, 
g0029368-173830, and NGC4395, which is a low-mass AGN with $\log M_{\mathrm{BH}}/M_\odot  = 3.96 \pm 0.22$ \citep{woo_ngc4395mass_2019} that will be discussed in Section \ref{subsubsec:ngc4395}, are shown in Figure \ref{fig:example_lc}. 

\subsection{Calibrating AGN parameters using photometry and lightcurves} \label{subsec:smss_calib}
\begin{figure*}
    \centering
    \includegraphics[width=\linewidth]{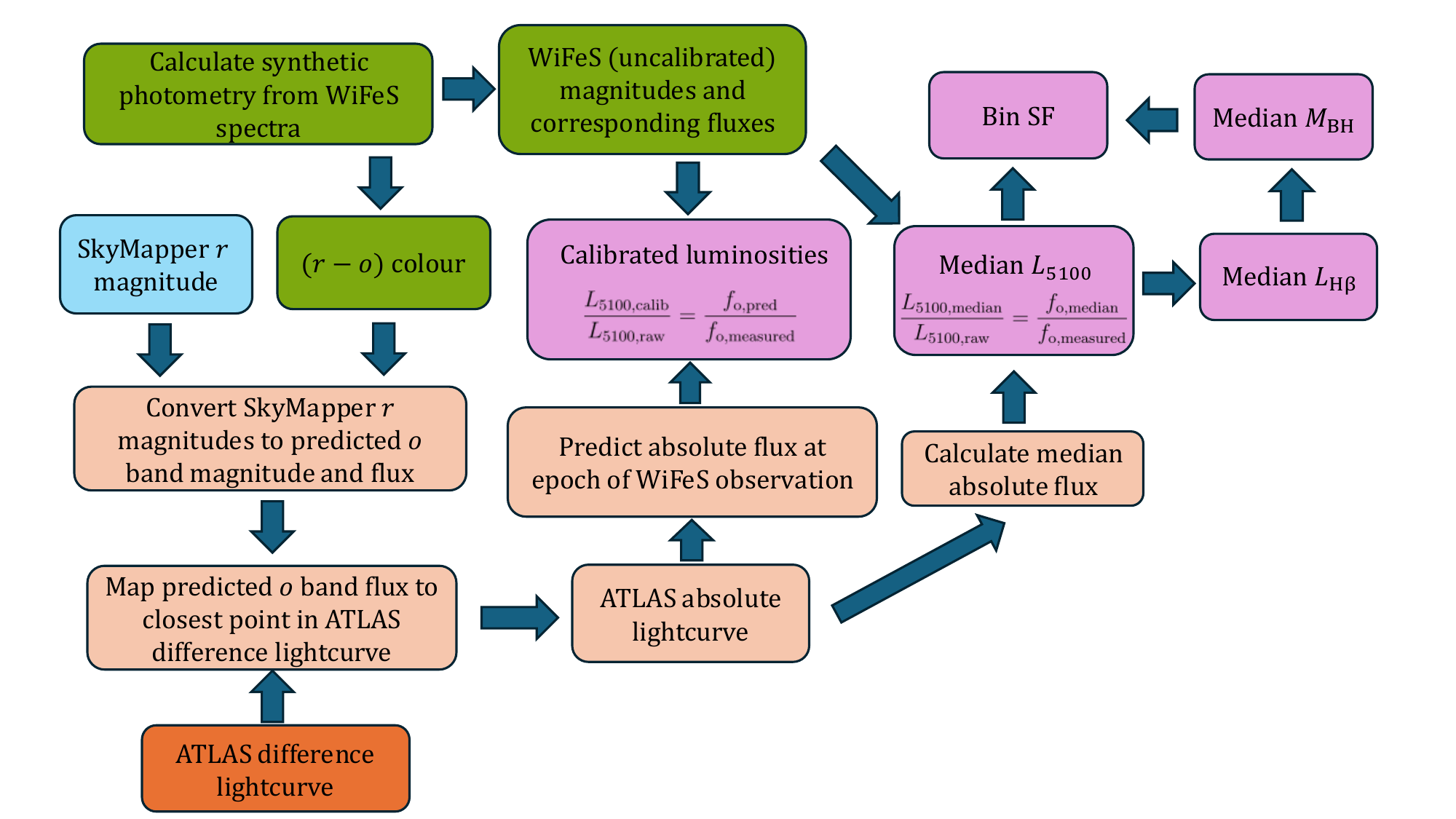}
    \caption{A summary of the calibration process described in Section \ref{subsec:smss_calib}.}
    \label{fig:flowchart}
\end{figure*}

One risk with using a single spectrum to measure AGN parameters such as the luminosity and black hole mass is that weather conditions could introduce errors if the target or the standard star were affected by clouds. The properties derived from a single spectrum are also not necessarily representative of the average AGN properties during the whole lightcurve. \refbftwo{\citet{amrutha_masscorrection_2026} have shown that over $\sim$20 years objects in this sample show RMS changes in luminosity of 0.26 dex and in mass of 0.47 dex. }Absolute photometry for all targets is available from SkyMapper but not for the exact epochs of the WiFeS spectra. The correction is provided by the ATLAS lightcurves which can predict absolute fluxes for WiFeS spectra. We can then use these absolute fluxes and the uncalibrated spectra, assuming constant spectral colours, to calculate calibrated $L_{5100}$ and $M_{\mathrm{BH}}$ measurements for the night of observation. Additionally, we can derive $L_{5100}$ and $M_{\mathrm{BH}}$ measurements that represent a median activity level by using the same method to predict absolute flux values corresponding to the median difference flux of the lightcurves. A summary of the process is shown in Figure \ref{fig:flowchart}.

The WiFeS spectra were extracted using a 6.7$^{\prime\prime}$ aperture to match the 6dFGS spectra for a spectral comparison \citep{amrutha_clagn_24,amrutha_masscorrection_2026}. We integrate the spectra over the ATLAS and SkyMapper filters to obtain synthetic photometry, which is converted to a flux $f_{\mathrm{o, measured}}$, as well as the $(r-o)$ and $(g-c)$ colours. Note, that the colour terms are small as the pivotal wavelengths are similar and there are significant overlaps between filters, with $\lambda_{\mathrm{pivot}, r} = 614 \ \mathrm{nm}$, $\lambda_{\mathrm{pivot}, o} = 675 \ \mathrm{nm}$, and $\lambda_{\mathrm{pivot}, g} = 508 \ \mathrm{nm}$, $\lambda_{\mathrm{pivot}, c} = 529 \ \mathrm{nm}$. To get well-calibrated absolute fluxes, we use 6$^{\prime\prime}$ aperture-corrected $g$ and $r$ magnitudes (\texttt{apc06}) from Data Release 4 of the SkyMapper Southern Sky Survey \citep{onken_skymapper_2024}. We only use photometry taken after MJD 57900, with \texttt{flags}<4, \texttt{nimaflags}<5, and \texttt{img\_qual}<3 as a quality cut. The 6$^{\prime\prime}$ aperture was chosen to be consistent with the 6.7$^{\prime\prime}$ aperture used for WiFeS spectra, as the \texttt{apc06} magnitudes have been corrected for SkyMapper PSF losses. To verify consistency, we compared the \texttt{apc06} magnitudes with synthetic photometry from the WiFeS spectra. The two measurements lie on the 1:1 relation with a scatter of $\sim 0.3$ mag, indicating that, on average, they are on the same scale. As a comparison, the 8$^{\prime\prime}$ (\texttt{apc08}) SkyMapper magnitudes are systematically brighter than the 6.7$^{\prime\prime}$ magnitudes from synthetic photometry. 

Using the spectral $(r-o)$ colour, we convert the SkyMapper $r$ absolute fluxes into a predicted $o$ band flux. These $o$ band absolute fluxes are mapped to difference fluxes closest to the date of the SkyMapper measurements. Using this mapping, we can estimate the absolute $o$ band flux at the WiFeS epochs, represented by $f_\mathrm{o,pred}$ in Figure \ref{fig:flowchart}. Since there are multiple SkyMapper measurements, we can repeat this process for each measurement to get multiple predicted fluxes at the WiFeS epoch and average them for a more precise measurement. 

To calibrate $L_{5100}$, we assume a constant colour, such that the ratio between the calibrated and measured $L_{5100}$ is equal to the ratio between the predicted $o$ band flux at the WiFeS epoch and the $o$ band flux from WiFeS synthetic photometry. The ATLAS lightcurves probe the changes in continuum but not the line luminosity, so we assume that the broad \hb \ line follows $\Delta \log L_{\mathrm{H\beta}} \approx \Delta \log L_{5100}$. This slope of $\sim$1 is observed in our sample and in other works such as by \citet{rakshit_stalin_2020}. We can apply a similar procedure to obtain AGN luminosities and masses corresponding to the average activity level during the lightcurve duration. Instead of calculating the predicted fluxes at the WiFeS epoch, we map the lightcurve median difference flux to an absolute flux, then use that flux to calibrate $L_{5100}$ and $L_{\mathrm{H\beta}}$. The calibrated $L_{\mathrm{H\beta}}$ is used to calculate a calibrated $M_{\mathrm{BH}}$, which is used with the calibrated $L_{5100}$ to define SF sample bins. 

\subsection{Subtracting the host component from lightcurves}
The median ATLAS-o and ATLAS-c magnitudes obtained by using SkyMapper magnitudes, ATLAS difference lightcurves, and WiFeS spectra as described in \ref{subsec:smss_calib} are used to shift the difference lightcurves to an absolute scale. We estimate the host galaxy flux by multiplying the host fraction by the median absolute flux, then subtract it from the absolute lightcurves to produce AGN-only absolute lightcurves. Individual host fraction estimates may be noisy from the degeneracy between host templates and blue AGN continua, as well as from the uncertainty in the host light captured by the ATLAS PSF. Thus, we use the median host fraction in each SF bin, weighted by the number of data points in each lightcurve, to perform the host correction for all objects in the bin. For seven objects, subtracting the median host fraction results in negative fluxes in the orange and/or cyan absolute lightcurve, so we remove them from our sample. These objects constitute less than 3\% of our sample and will not impact the overall findings.

\section{Structure Function Methods}\label{sec:sf_methods}
In this section, we describe the methods we use to characterise optical variability. We also discuss the process of estimating the noise in the SF. Next, we describe the process of fitting the slope, break timescales, and amplitude of the SF. 

\subsection{Calculating the SF} \label{subsec:method_sf}
The definition of the structure function used in this work is 
\begin{equation}
    A (\Delta t) = \sqrt{k^2(\langle \Delta m \rangle_{\mathrm{median}} ^2 - \langle \sigma^2 \rangle_{\mathrm{median}})},
\end{equation} \label{eq:sf_measured}
similar to that used by \citet{di_clemente_variability_1996}, though we elect to use the median instead of the mean to minimise the effect of outliers. $A$ is the variability amplitude, $\Delta m$ is the magnitude difference between two measurements, $\sigma$ represents the error, and $k \approx 1.4826$ is a constant to convert the median absolute deviation to a standard deviation. \refbf{$\Delta t$ is in the rest-frame, where $\Delta t = \Delta t_{\mathrm{obs}}/(1+z)$.}  In the $\Delta t$ axis, we create bins that contain an approximately equal number of data points. We only sample $\Delta t$ up to 50\% of the length of the lightcurve segments (MJD 57900-MJD 58882, MJD 58882-MJD 60815) since higher $\Delta t$ data will be plagued by window effects. To increase the signal-to-noise ratio of our structure functions, we construct ensemble structure functions by binning the sample by BH mass and AGN luminosity. When binning by a single parameter, we select bins so that each bin has approximately the same number of objects. In subsequent analyses of SF trends with both mass and luminosity, we adopt an alternative binning approach using overlapping bins to improve the sampling for fitting purposes. For each object, we define a bin that includes itself and the nine nearest neighbours in both luminosity and mass. The bins are constructed by iteratively and alternately expanding the bin width in luminosity and then in mass, in very small increments, until exactly ten objects are included.
\begin{figure}
    \centering
    \includegraphics[width=\linewidth]{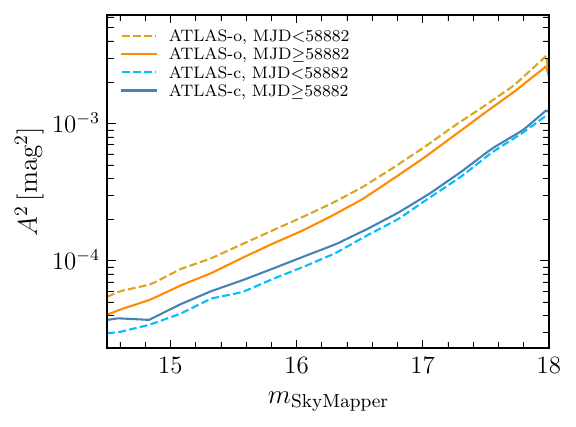}
    \caption{The error distribution with SkyMapper $r$ and $g$ magnitudes. The error for the ATLAS-o lightcurves are estimated by binning the star-forming galaxies in bins of SkyMapper $r$ magnitudes, and likewise for ATLAS-c and SkyMapper $g$ magnitudes. The errors before and after MJD 58882 are estimated separately because they appear to differ significantly. }
    \label{fig:err_mag}
\end{figure}

The error $\sigma$ is estimated using \refbf{difference lightcurves of} star-forming galaxies from 6dFGS with similar redshift and apparent magnitude as the AGN sample. We do not expect them to be variable, so any apparent variability can be attributed to noise. \refbf{For these galaxies, the fractional amplitude used can be directly calculated using the SkyMapper magnitudes without a host correction.} We assume that the error behaviour of SF galaxies and AGN with similar $g$ and $r$ magnitudes are similar because we do not have ATLAS-o and ATLAS-c magnitudes in a consistent 6.7" aperture for any of the galaxies. We sample the variability using the entire lightcurve in 20 SkyMapper magnitude bins and use linear interpolation to obtain the errors for each object. At $m_r, m_g<14$ and $m_r, m_g>18$, the error values are poorly sampled, so they are instead estimated from a curve fit to the rest of the points. The interpolation is shown in Figure \ref{fig:err_mag}. Note, that there are only three objects where $m_r>18$ and only seven objects where $m_g>18$. We find a noticeable difference in the error levels before and after the wallpaper change at MJD 58882, so we estimate these errors separately. 

\subsection{Measuring the slope and break timescale} \label{subsec:slope_method}
We do not assume a priori an agreement with the DRW model or a constant SF slope to avoid biasing our results. Instead, for each ensemble SF, we measure the slope of the region commonly assumed to be the random walk regime of the SF by fitting a power law over pair separation times of $\Delta t=14-100$ days using a least squares fit. We avoid time intervals shorter than 14 days to exclude points that may be affected by improper noise subtraction and to avoid fitting a possible short-term break, which has been found by a few studies \citep[e.g.][]{edelson_kepler_2014} on timescales of $\sim1-10$ days. The upper limit of 100 days was informed by visually inspecting the SFs, as any breaks appear to occur after 100 days, if at all. 

We model an SF with a break using a broken power law. However, breaks are likely not ubiquitous, so we use the Bayesian Information Criterion (BIC) to determine whether a broken power law or a single power law is the preferred model. After the decorrelation timescale, the SF slope is expected to be zero, so we fix the slope of the broken power law to that value. In addition, window effects may bias the fitted slope, so we do not allow a free slope at $\Delta t > \tau_{\mathrm{break}}$. It is generally recommended that the fitted break timescales should be less than 10\% of the baseline \citep{kozlowski_limitations_2017}, which corresponds to a maximum allowed break timescale of 190 days for our sample, though we allow the algorithm to return a best fit timescale of up to 200 days. All measured break timescales $\geq 190$ days will be considered as lower limits. Window effects may be interpreted by the fitting algorithm as breaks, so we only fit data up to a $\Delta t$ of 500 days. 

\subsection{Fitting the amplitude dependence on AGN properties}\label{subsec:method_amplitude_fit}
We measure the variability amplitude at 50 days, $A_{\mathrm{50d}}$, to study its dependence on $L_{5100}$ and $M_{\mathrm{BH}}$. To reduce noise from possible outliers at the selected $\Delta t$, we use the single power law fits to the random walk regime to get the best fit variability amplitude at $\Delta t = 50 \  \mathrm{d}$. Then, we evaluate the dependence of the amplitude on AGN parameters by fitting the following equation to the data:
\begin{equation} \label{eq:amp_dep}
    \log A (50 \, \mathrm{ d}) = a M_7 + b L_{43} + c, 
\end{equation}
where $M_7 = \log (M_{\mathrm{BH}}/10^7M_\odot)$, and $L_{43} = \log (L_{5100}/10^{43}\mathrm{erg \ s^{-1}})$. The statistical measurement uncertainties of the luminosity and BH mass from spectral decomposition are insignificant compared to the systematic uncertainties from the viewing angle, anisotropy, and single-epoch black hole mass measurement, so we do not use them as errors in the fitting process and instead assume equal weights. We do not attempt to fit a wavelength dependence using the cyan and orange data, as the variability in different wavelengths may not have the same dependence on BH mass and luminosity. 

\subsection{Extending the baseline using two-epoch spectra}
\label{subsec:two_epoch}
For most of the sample, we have 6dFGS spectra from $\sim 20$ years ago \citep{amrutha_masscorrection_2026}, which can be used to calculate the change in AGN luminosity across 20 years, and inform us about whether the SFs continue to rise on timescales longer than those we probe. A detailed description of the 6dFGS spectra fitting process is described by \citet{amrutha_masscorrection_2026}. Due to the lower spectral resolution of the 6dFGS spectra, the uncertainty in the AGN-host decomposition is likely higher and will reduce the precision of $L_{5100}$ measurements. Thus, we use the change in the broad H$\beta$ luminosity as a proxy for the change in $L_{5100}$ by approximating $\Delta \log L_{\mathrm{H\beta}} \approx \Delta \log L_{5100}$, which is similar to the slopes found in the literature \citep[e.g.][]{rakshit_stalin_2020} and the correlation within the sample. The SF of the two-epoch data is given by Equation \ref{eq:2_epoch}, 
\begin{equation} \label{eq:2_epoch}
    A_{\mathrm{H\beta}} (\Delta t) = \sqrt{k^2(\Delta m _{\mathrm{H\beta}} ^2 - \langle \sigma_{\Delta m_{\mathrm{H\beta}}}^2 \rangle_{\mathrm{median}})}, 
\end{equation}
where $\Delta m_{\mathrm{H\beta}}$ is given by Equation \ref{eq:2_epoch_deltam}, 
\begin{equation} \label{eq:2_epoch_deltam}
    \Delta m_{\mathrm{H\beta}} =  |2.5\log(\frac{L_{\mathrm{H\beta, 6dFGS}}}{L_{\mathrm{H\beta, WiFeS}}})|.
\end{equation}
$L_{\mathrm{H\beta, 6dFGS}}$ and $L_{\mathrm{H\beta, WiFeS}}$ are the H$\beta$ luminosities measured from 6dFGS and WiFeS spectra, $k \approx 1.4826$, and $\sigma_{\Delta m_{\mathrm{H\beta}}}$ is the error in $\Delta m_{\mathrm{H\beta}}$ propagated from measurement uncertainties. There are likely differences in the flux profile captured by the ATLAS PSF, 6dFGS spectra, and WiFeS spectra, but since the SF uses the relative brightness, this effect should not bias either the ATLAS or 6dFGS/WiFeS SFs.

\section{Results}\label{sec:discussion}
\subsection{Overall SF behaviour and slopes}\label{subsec:slope}
\begin{figure*}
    \centering
    \includegraphics[width=\linewidth]{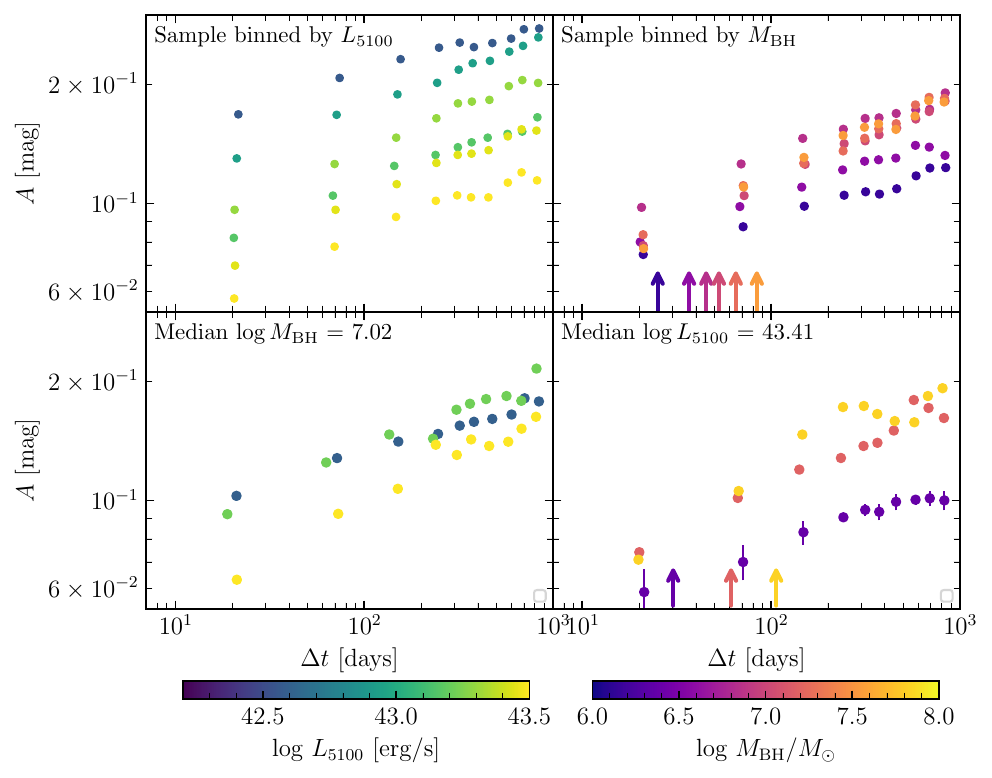}
    \caption{The top left and right panels show the SFs binned by only luminosity and only BH mass respectively. The bottom panels show SFs binned by both luminosity and BH mass. In the bottom left panel, all three SFs have similar BH mass and different luminosities, and vice versa for the bottom right panel. In the right panels, the break timescales predicted by \citetalias{burke_characteristic_2021} are indicated with arrows.}
    \label{fig:vsf_binmasslum}
\end{figure*}

We first bin the SFs by only BH mass and luminosity separately to demonstrate the overall trends, as shown in the top panels of Figure \ref{fig:vsf_binmasslum}. In the top left panel, there is a clear inverse dependence of variability amplitude \refbftwo{on} $L_{5100}$. In the top right panel, we observe a positive correlation of the amplitude with BH mass with some scatter, which is not unexpected since the typical BH mass uncertainty is $\sim 0.5 \ $dex \citep{shen_mass_2013}. We also overplot the break timescales predicted using the $\tau_{\mathrm{break}}-M_{\mathrm{BH}}$ relation in \citetalias{burke_characteristic_2021} as arrows, and note that we see no signs of breaks at the predicted timescales. While the low mass bins show signs of a turnover, the high mass SFs appear to keep rising even at $\Delta t\sim 1000 \ \mathrm{d}$. In both panels, there appears to be a dependence of the slope on luminosity and BH mass. We also see wave-like features in the SF at high $\Delta t$, which indicate window effects, so the apparent breaks may be artefacts. 

Only binning the SFs by luminosity may smooth out features dependent on BH mass and vice versa, so we also bin the SF by both parameters simultaneously. As described in Section \ref{subsec:method_sf}, to avoid bias from fixed bin widths and to ensure uniform weighting across the entire parameter space, we construct a bin around each object that contains nine other objects closest in luminosity and BH mass. In the bottom panels of Figure \ref{fig:vsf_binmasslum}, we present six binned SFs. Three bins with similar BH mass but differing in $L_{5100}$ are presented in the bottom left panel, and three bins with similar BH mass but different $L_{5100}$ are presented in the bottom right panel. Similarly, we see that the variability amplitude is anticorrelated with the luminosity and positively correlated with the BH mass, and again observe no breaks at the timescales predicted by \citetalias{burke_characteristic_2021}. The slope-BH mass trend is more prominent compared to the top right panel, suggesting that while only binning by mass could yield a higher signal to noise, the blending of different regimes in one bin will dilute features in the SF. Consequently, we adopt this binning method in both mass and luminosity for the remainder of this paper. 

\begin{figure*}
    \centering
    \includegraphics[width=\linewidth]{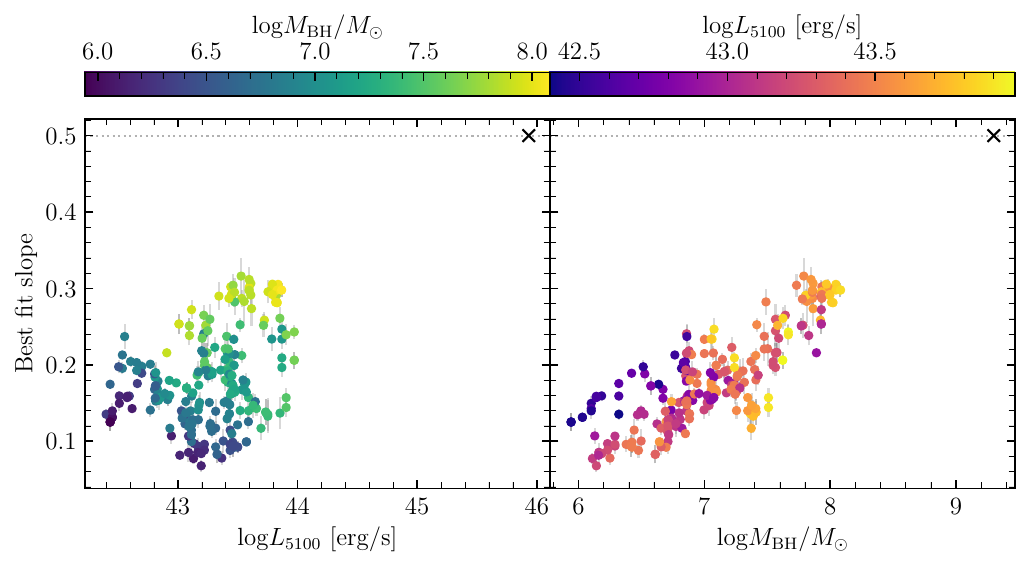}
    \caption{\refbf{Ensemble SF slopes, using data at 14 days< $\Delta t$< 100 days, vs. luminosity (left) and BH mass (right), compared to results from \citet{tang_universality_2023} (crosses, representing their sample medians). Our slopes are below 0.5 (DRW, dotted line), increase with mass, and appear independent of luminosity.}}
    \label{fig:slope_mass_lum}
\end{figure*}

The measured slopes of the binned SFs plotted against $L_{5100}$ and BH mass are shown in Figure \ref{fig:slope_mass_lum}. In the left panel, there appears to be no trend of the slope with $L_{5100}$, whereas in the right panel, there is an obvious trend of the slope increasing with BH mass, up to a slope of $\sim0.3$ at the high mass end. \refbf{The average slope found by \citet{tang_universality_2023} in high-luminosity quasars 
is also plotted at its median $L_{5100}$ \citep[][estimated from $L_{3000}$ with a bolometric correction from \citealp{richards_spectral_2006}]{rakshit_stalin_2020} and median $M_\mathrm{BH}$. If our slope-mass relation persists to high $M_\mathrm{BH}$, it will be consistent with the \citet{tang_universality_2023} 
results.} 
Fitting a dependence of the slope on these parameters, we obtain, 
\begin{equation} \label{eq:slope_fit}
    \gamma = (0.12 \pm 0.01) M_7-(0.04 \pm 0.01)L_{43}+(0.19\pm 0.01).
\end{equation}
Equation \ref{eq:slope_fit} implies little luminosity dependence and a significant BH mass dependence. To verify whether the slope dependence changes with mass or luminosity, we also fit a mass dependence in luminosity bins and vice versa. We find no slope-luminosity dependence in all mass regimes, and a positive slope-BH mass dependence in all luminosity regimes. Following the convention in some literature, we also fit the slope dependence on the BH mass and the Eddington ratio. Note, that the Eddington ratio is estimated from the BH mass and AGN luminosity, so it would be difficult to disentangle effects of the BH mass, luminosity, and accretion rate/Eddington ratio on the variability. In addition, BH mass measurements have an uncertainty of $\sim$0.5 dex, which adds another dimension of uncertainty to the $M_{\mathrm{BH}}$ and $R_{\mathrm{Edd}}$ parametrisation. Therefore, for the majority of the paper, we discuss the correlations with the luminosity and BH mass instead of the more uncertain \refbftwo{Eddington ratio and BH mass.} For completeness, the best fit equation \refbftwo{in the latter case} is,
\begin{equation}
    \gamma = (\refbftwo{0.08} \pm 0.01)M_7-(\refbftwo{0.03} \pm 0.01) \log R_{\mathrm{Edd}}+(\refbftwo{0.15}\pm 0.01).
\end{equation}
The degeneracy between $L_{5100}$, $M_{\mathrm{BH}}$, and $R_{\mathrm{Edd}}$ is evident when comparing the best fit coefficients, where the fit coefficient for $\log R_{\mathrm{Edd}}$ and $\log M_{\mathrm{BH}}$ are consistent with the fit coefficients for $\log L_{5100}$ and $\log M_{\mathrm{BH}}$, as $R_{\mathrm{Edd}} \propto L_{5100}/M_{\mathrm{BH}}$. 

We consider whether any systematic effects could produce the significant trend of the slope with BH mass and the deviation from $\gamma=0.5$.  Mixing objects with different break timescales within an ensemble SF bin could in principle result in slopes shallower than 0.5. However, this is unlikely to generate a mass-dependent trend, since objects were binned with other objects with a similar BH mass and luminosity. While the over- or under-subtraction of magnitude errors could affect the slope, the errors are correlated with the apparent magnitude, which has no significant correlation with the BH mass due to the presence of host galaxy emission. Thus, this correlation cannot be attributed to misestimation of error. 

\subsection{Breaks in the SF} \label{subsec:result_breaks}
\begin{figure}
    \centering
    \includegraphics[width=\linewidth]{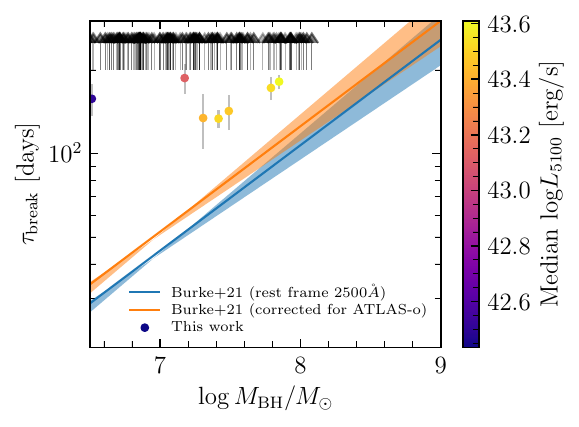}
    \caption{A comparison of the fitted break timescales in bins of mass and luminosity compared to the scaling relation found by \citetalias{burke_characteristic_2021}. The lower limits are indicated by arrows. Contrary to the predicted relation, we see no correlation of the break timescale with BH mass. }
    \label{fig:break_vs_mass}
\end{figure}

\citetalias{burke_characteristic_2021} suggested that break timescales scale with BH mass, as $\tau_{\mathrm{break}} = 107^{+11}_{-12} \mathrm{days} (M_{\mathrm{BH}}/10^8 M_{\odot})^{0.38^{+0.05}_{-0.04}}$. For our mass range from $\log M_{\mathrm{BH}}/M_\odot = 5.5$ to $\log M_{\mathrm{BH}}/M_\odot = 8.6$, we would be able to measure the breaks with our $\sim2000$ day-long lightcurves. As described in Section \ref{subsec:slope} and shown in the right panels of Figure \ref{fig:vsf_binmasslum}, we do not see any signs of breaks on these timescales. \citetalias{burke_characteristic_2021} associate the break timescale with the thermal or orbital timescale, which according to a Keplerian thin disc model would follow a scaling relation of $t_{\mathrm{orb}} \propto L^{0.75}M^{-0.5}$, which they suggest to be consistent with their measured slope within 2.5$\sigma$. We measure the break timescales of ensemble SFs binned by luminosity and BH mass, and compare them with the \citetalias{burke_characteristic_2021} relation in Figure \ref{fig:break_vs_mass}. \citetalias{burke_characteristic_2021} used a dataset with a range of rest-frame wavelengths, and used the relation $\tau \propto \lambda^{0.17}$ found by \citet{macleod_modeling_2010} to scale the timescale to a timescale at a rest-frame wavelength of 2500\AA. For consistency, we use the same relation to obtain predicted timescales by \citetalias{burke_characteristic_2021} at a rest-frame wavelength of $\sim6500$\AA \ to compare them to our measurements in ATLAS-o. As indicated by the black arrows in Figure \ref{fig:break_vs_mass}, most of our measurements are lower limits, so the few data points with valid measurements cannot provide meaningful constraints on the $\tau_{\mathrm{break}}-M_{\mathrm{BH}}-L_{5100}$ correlation. 
It is likely that our lightcurves and those used by \citetalias{burke_characteristic_2021} are insufficiently long relative to the true decorrelation timescale. A common recommendation is that the break timescales should be less than 10$\%$ of the baseline \citep{kozlowski_limitations_2017}. While \citetalias{burke_characteristic_2021} implemented this recommendation by discarding fits that do not meet this criteria, this could have caused a selection effect where the light curve length is the driver behind their measured relation. The correction of $\tau \propto \lambda^{0.17}$ could also have affected the results by \citetalias{burke_characteristic_2021}, since other works have found different scaling relations \citep{stone_optical_2022,stone_correction_2023}. In the following subsections, we discuss the possibility of multiple breaks in the SF, and use spectra taken 20 years apart to infer whether the SFs keep rising on decadal timescales. 

\subsubsection{A short-term break in NGC4395} \label{subsubsec:ngc4395}
\begin{figure}
    \centering
    \includegraphics[width=\linewidth]{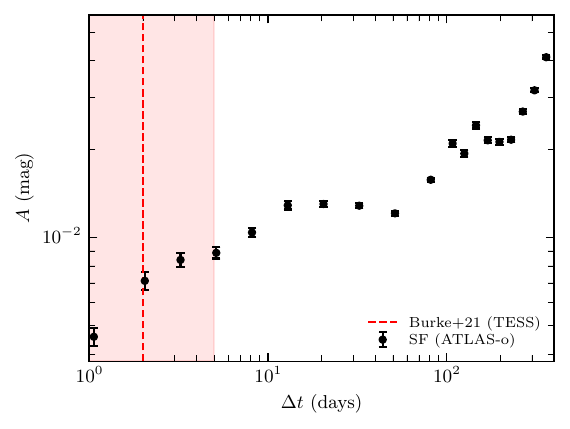}
    \caption{\refbf{ATLAS-o band variability structure function of NGC4395. There are signs of a short-term break into a flat SF,} but the SF rises again after $\sim$ 50 days. This SF is not corrected for host emission, which will affect the amplitude but not the overall shape. The break timescale from \citetalias{burke_characteristic_2021} and its uncertainty are shown as a dashed line and shaded region. The cadence of ATLAS lightcurves is likely not high enough to probe such short-time breaks.}
    \label{fig:ngc4395}
\end{figure}

Some studies in low-luminosity AGN have suggested the presence of short-term breaks on the scale of $\sim 1-10$ days \citep{edelson_kepler_2014,kelly_flexible_2014}. For example, \cite{edelson_kepler_2014} found a break timescale at $\sim 5$ days for Zw 229-15, and suggested that there will be another break at a longer timescale corresponding to the viscous timescale of the accretion disc. Short-term breaks could have a different origin from any long-term breaks, but it may be difficult to distinguish between the two in the absence of a sufficiently long baseline to probe the rise of the SF after a short-term break. Therefore, some of the break timescales measured in the literature could be short-term breaks instead of the long-term breaks that are sometimes associated with orbital/thermal timescales. For example, one object from the \citetalias{burke_characteristic_2021} sample, NGC 4395, has a BH mass of $\log M_{\mathrm{BH}}/M_\odot  = 3.96 \pm 0.22$ \citep{woo_ngc4395mass_2019} and a measured break timescale of $\sim 2$ days. We show the ATLAS lightcurve of NGC4395 in the bottom panel of Figure \ref{fig:example_lc}, and its SF in Figure \ref{fig:ngc4395}. The ATLAS lightcurve is likely not well-sampled enough to probe the quoted break timescale accurately, but we observe hints of a flattening in the structure function around $\sim 10$ days. However, the SF rises again at $\Delta t \sim 50$ days, so \citetalias{burke_characteristic_2021} could have measured a short-term break in this object. It is unclear whether short-term breaks are common, \refbf{%
what their 
physical meaning could be, and whether they are sampling artefacts
(see Section \ref{subsec:result_two_epoch} for a general discussion of window/sampling effects). 
If they are indeed physical, }we speculate that they could be caused by two variability processes, one low-amplitude process that saturates on short timescales and another high-amplitude, long-timescale process. While our ensemble SFs do not show signs of a short-term break, we inspect the SFs of individual objects to verify if this could be affecting our results. A small fraction of our sample display signs of a short-term break and subsequent rise, but these objects are rare enough that they will likely not drive the behaviour of the ensemble SFs. 

\subsubsection{Extending the baseline to 20 years} \label{subsec:result_two_epoch}
\begin{figure*}
    \centering
    \includegraphics[width=\linewidth]{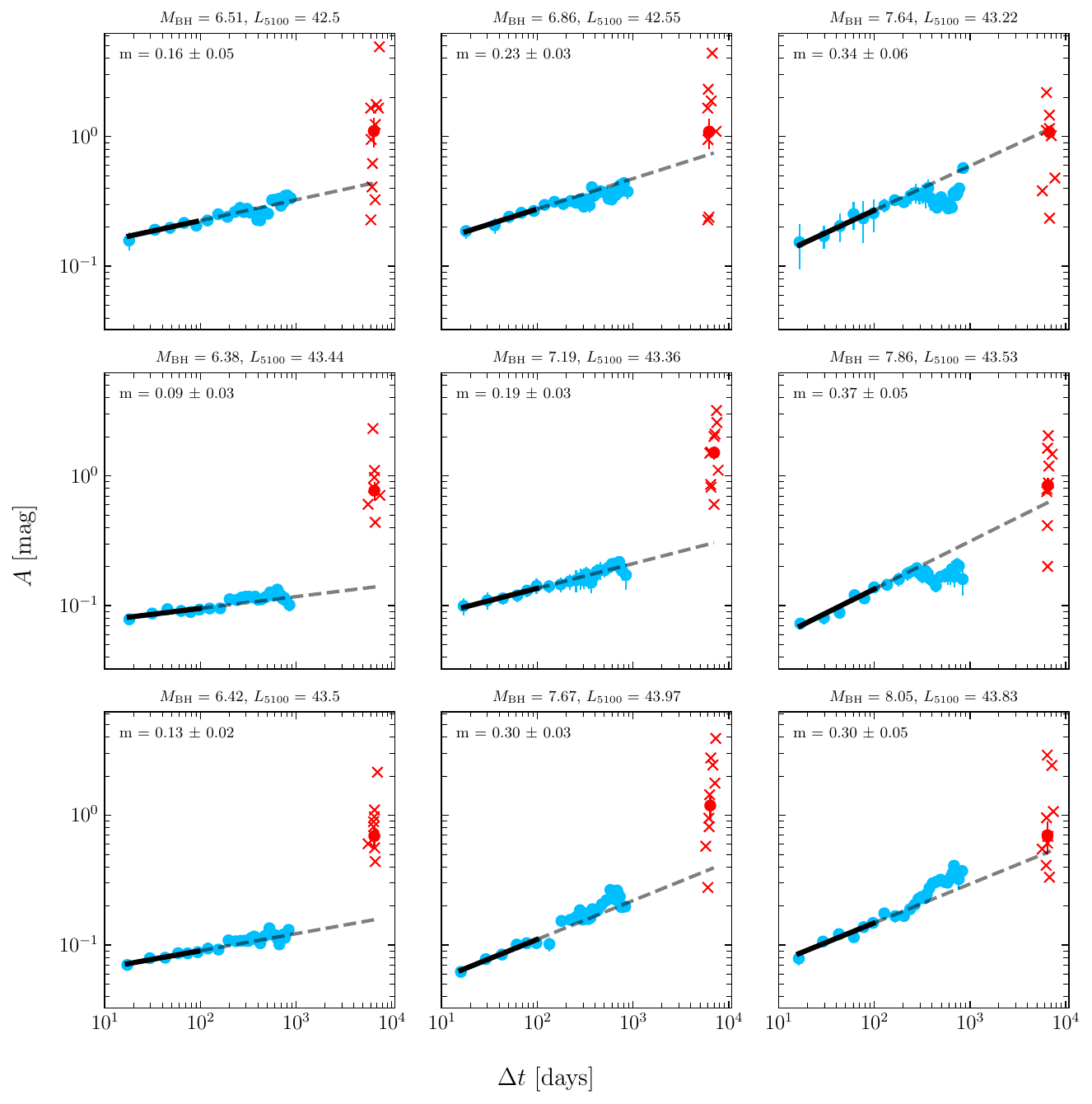}
    \caption{\refbf{Variability structure functions in the cyan band binned by mass and luminosity. A power law fit (solid line) is extrapolated to $\Delta t \sim$20~years (dashed line). Additionally, variability of broad emission lines from two-epoch $\mathrm{H\beta}$ observations is shown as crosses and their median as filled circles.}}
    \label{fig:vsf_panel}
\end{figure*}

The validity of interpreting the break timescales has also been questioned by authors such as \citet{emmanoulopoulos_use_2010}, which demonstrated that breaks can be recovered from simulated lightcurves generated from a featureless intrinsic PSD, and \citet{bauer_quasar_2009} suggested that the high $\Delta t$ end would have underestimated amplitudes due to sampling effects. \citet{stone_optical_2022} also found that the recovered break timescales increase with lightcurve baselines, suggesting that timescale measurements in the literature are affected by window effects. While we do not have baselines as long as \citet{stone_optical_2022}, we can use spectroscopic data taken $\sim 20$ years apart to check whether the SF keeps rising after the apparent break we observe in our sample. In Figure \ref{fig:vsf_panel}, we present nine ensemble structure functions with different median masses and luminosities, plotted with two-epoch SFs calculated from 6dFGS and recent WiFeS spectra. Since the continuum luminosity measurements may be affected by inaccurate host estimates, we use $\Delta L_{\mathrm{H\beta}}$ as a proxy for $\Delta L_{5100}$, and compare the two-epoch SF with the ATLAS-c SF, which has a pivot wavelength of $\lambda_{\mathrm{pivot}, c} = 529 \ \mathrm{nm}$. In our sample and in a number of studies \citep[e.g.][]{rakshit_stalin_2020}, it has been observed that $\Delta L_{5100} \sim \Delta L_{\mathrm{H\beta}}$. The blue points represent our cyan SFs, while the red crosses denote the two-epoch data. The median SFs of the two-epoch data are shown as red points. The SFs continue to rise on timescales of $\sim 20$ years even after the apparent flattening at the long edge of our SFs. Therefore, we conclude that the apparent breaks in our sample are likely products of window effects. If there are long-term breaks in low-luminosity AGN, the timescales are too long to be probed with our $\sim 2000$ day baseline. 

\subsection{Dependence of variability amplitude on AGN parameters} \label{subsec:amplitude_fit}
The best fit parameters to Equation \ref{eq:amp_dep} at $\Delta t=50$d in the orange and cyan band are 
\begin{eqnarray} \label{eq:bf_amp_dep_o}
    \log A_{\mathrm{o, 50d}} & = & - 0.81 + 0.15 M_7 -0.39 L_{43} \\
\label{eq:bf_amp_dep_c}
    \log A_{\mathrm{c, 50d}} & = & - 0.74 + 0.16 M_7 - 0.38 L_{43} \\
\label{eq:bf_amp_dep_redd_o}
    \log A_{\mathrm{o, 50d}} & = & - 1.23 - 0.22 M_7 -0.37 \log R_{\mathrm{Edd}}  \\
\label{eq:bf_amp_dep_redd_c}
    \log A_{\mathrm{c, 50d}} & = & - 1.16 -0.21 M_7-0.37 \log R_{\mathrm{Edd}}  ~. 
\end{eqnarray}
The measurement uncertainty of the best fit parameters is  0.01\refbf{, and the scatter of the residuals is $\sigma \sim 0.05$}. We find an anticorrelation of the variability amplitude with $L_{5100}$ and a positive correlation with $M_{\mathrm{BH}}$ in both filters. \refbf{Plotting} the residuals of the fit against $L_{5100}$ and BH mass\refbf{, we find} no obvious trend of the residuals with both parameters, suggesting that the adopted functional form does not bias one region of the parameter space. This is contrary to some results in the literature that suggest an anticorrelation with BH mass at low accretion rates \citep{smith_kepler_2018,yu_examining_2022}. The best fit coefficients of the orange and cyan band are also statistically consistent with one another. Again, for completeness, we fit a dependence on the BH mass and Eddington ratio, noting the difficulty in disentangling effects of $L_{5100}$, $M_{\mathrm{BH}}$, and $R_{\mathrm{Edd}}$ as discussed in Section \ref{subsec:slope}. The best fit parameters are presented in Equation \ref{eq:bf_amp_dep_redd_o} and Equation \ref{eq:bf_amp_dep_redd_c}. The variability amplitude is not solely dependent on the Eddington ratio, so the variability is likely not mainly driven by accretion rate fluctuations. 

Note, that the best fit parameters will change if we fit a different $\Delta t$ due to the varying SF slopes on the sample. If we fold in Equation \ref{eq:slope_fit} from Section \ref{subsec:slope}, we obtain,
\begin{multline}
    \log A = - (1.13\pm0.02) -(0.39\pm0.01)L_{43}-(0.05\pm0.02)M_7 \\
    + ((0.12\pm0.01)M_7+(0.19\pm0.01))\log\Delta t ,
\end{multline}
suggesting that the amplitude-mass dependence vanishes at $\Delta t \sim 2.6$ days. In the literature, studies typically find an inverse correlation of the variability amplitude with AGN luminosity, while the correlation with BH mass is less consistent across studies. Due to different SF definitions, it is not straightforward to compare the values of the best fit parameters with the literature. Many studies associate the dependence on luminosity and mass to disc timescales, which could affect the observed variability. For example, the thermal timescale dictates the time needed for local thermal fluctuations to affect the entire disc. The measured amplitude correlations have also been associated with the orbital or thermal timescale of the accretion disc. \citet{tang_universality_2023} found that their best fit parameters correspond to the thermal timescale of the accretion disc, and that scaling the time axis of the SF with the thermal timescale produced a universal structure function that well-describes the sample regardless of their fundamental properties. This result suggests that the variability they probe grows with the orbital or thermal timescale. Since these timescales scale with the luminosity and BH mass, each SF bin will roughly have the same timescale. Thus, due to the variety of measured SF slopes, it is apparent that scaling the time axis will not result in a similar universal structure function. The difference from \citet{tang_universality_2023} may arise from the fact that we probe a different region of the accretion disc, which may be dominated by a different variability mechanism.

\section{Seyfert types 1.0 to 1.9: accretion variation or obscuration sequence?} \label{sec:extinction}
Type 1 Seyferts are classified into subtypes of 1.0, 1.2, 1.5, 1.8 and 1.9 based on the strength of the \refbftwo{broad} \hb \ line relative to the narrow [OIII] line \citep{winkler_1992}. The range in observed line ratios is often attributed to dust extinction {\citep[e.g.][]{maiolino_low-luminosity_1995} or variations in the accretion state \citep[e.g.][]{burtscher_davies_2016}. Given its large spatial extent, the Narrow Line region (NLR) \citep{capetti_seyfertnlr_1996,dopita_s7_2015} is not affected by nuclear obscuration or short-term luminosity variations.} Our almost volume-complete sample offers an opportunity to \refbf{investigate whether dust extinction is solely responsible for the diversity in $\mathrm{H\beta/[OIII]}$ line ratios by considering the colour of the variable emission component.}

\begin{figure*}
    \centering
    \includegraphics[width=\linewidth]{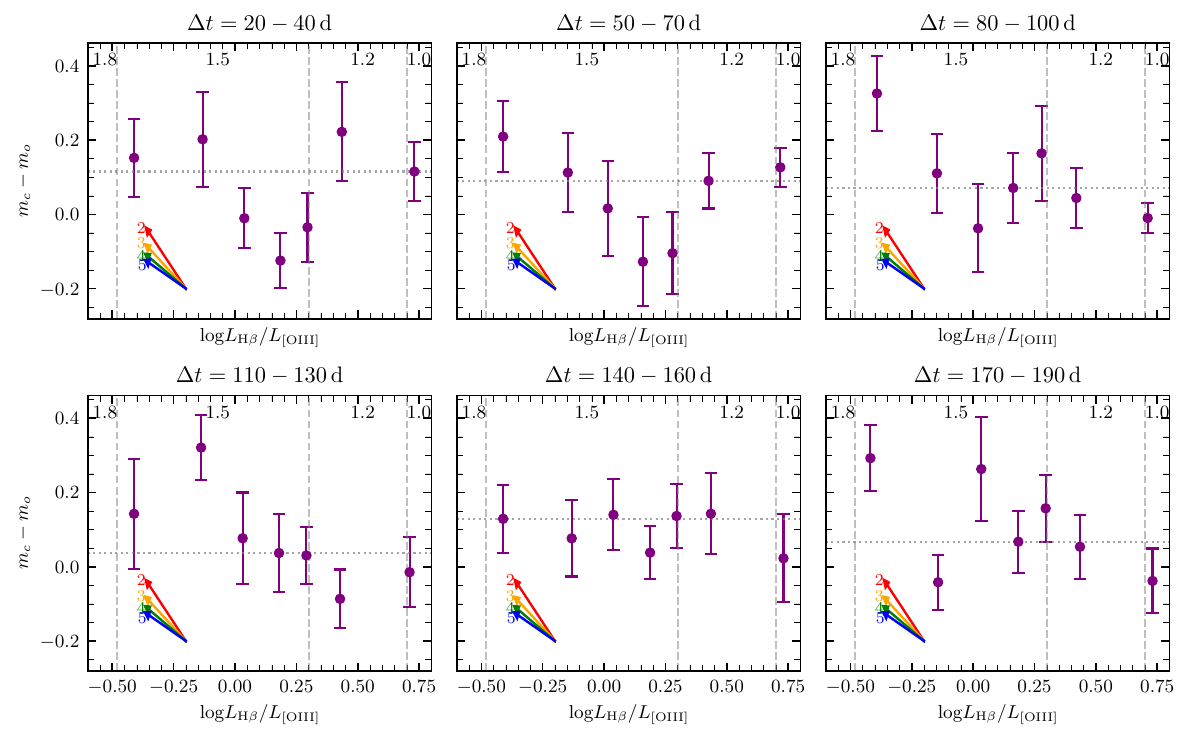}
    \caption{The median colour of the variable emission component $m_c-m_o$ plotted against the median $\log L_{\mathrm{H\beta}}/L_{\mathrm{[OIII]}}$ at $\Delta t$ centred at 30, 60, and 90, 120, 150, and 180 days. The arrows indicate the extinction slope derived from the \citet{calzetti_dust_2000} curve with an $R_V$ of 2, 3, 4, and 5. The vertical dashed lines indicate the divisions between Seyfert subtypes\refbf{, and the horizontal dotted lines indicate the median $m_c-m_o$.}}
    \label{fig:cmo_ratio}
\end{figure*}
\refbf{According to common extinction laws \citep[e.g.][]{cardelli_ext_1989,calzetti_dust_2000}, interstellar dust will extinguish the variable flux more strongly at shorter wavelengths, although some authors speculate about nuclear dust in AGN producing grey extinction \citep[e.g.][]{maiolino_dust_2001}. If the variety in $L_{\mathrm{H\beta}}/L_{\mathrm{[OIII]}}$ ratio is driven by dust extinction, then AGN with relatively weaker H$\beta$ lines will show redder variable flux components.} 

Here, we define the colour of the variable component as $m_c-m_o = -2.5 \log (\langle \Delta f_c \rangle/ \langle \Delta f_o \rangle)$, where $\langle \Delta f_c \rangle$, $\langle \Delta f_o \rangle$ are the cyan and orange median SF in flux units. $m_c-m_o$ may not be independent of the sampled $\Delta t$, so we sample pairs \refbf{in several bins of time separation,} 20-40 days, 50-70 days, 80-100 days, 110-130 days, 140-160 days, and 170-190 days, avoiding higher $\Delta t$ due to possible window effects. For each object, we first compute the median $m_c-m_o$, then we calculate the median $m_c-m_o$ in bins of $\log L_{\mathrm{H\beta}}/L_{\mathrm{[OIII]}}$. The line ratio bins were chosen to contain approximately equal numbers of objects, and we exclude objects with variability amplitudes that do not exceed the measurement uncertainty. 

For comparison, we derive an expression representing how the colour of the variable emission component changes with the line ratio if it were due to dust extinction, 
\begin{equation} \label{eq:co_colour}
    \frac{\Delta(m_c-m_o)}{\Delta (\log L_{\mathrm{H\beta}}/L_{\mathrm{[OIII]}})} = -2.5\frac{\kappa(c)-\kappa(o)}{\kappa(\mathrm{H\beta})}. 
\end{equation}
We obtain $\kappa(c)$, $\kappa(o)$, and $\kappa(\mathrm{H\beta})$ from the \citet{calzetti_dust_2000} extinction curve using an $R_V$ of 2, 3, 4, and 5. Figure \ref{fig:cmo_ratio} shows the median $m_c-m_o$ plotted against the median $\log L_{\mathrm{H\beta}}/L_{\mathrm{[OIII]}}$, sampling time intervals centred on 30, 60, 90, 120, 150, and 180 days. Arrows representing extinction vectors from Equation \ref{eq:co_colour} with a range of $R_V$ are plotted. We see a range of behaviour in all panels, from a slight decrease of $m_c-m_o$ with $\log L_{\mathrm{H\beta}}/L_{\mathrm{[OIII]}}$ to no obvious trend. If the Seyfert subtypes originate from varying degrees of dust extinction, we expect the points to be consistent with the extinction arrows. Given that $\log L_{\mathrm{H\beta}}/L_{\mathrm{[OIII]}}$ changes on timescales longer than what we probe \citep{amrutha_masscorrection_2026}, the different behaviour at different $\Delta t$ is likely a product of noise, and the Seyfert subtypes are not an extinction sequence. 

\refbf{In principle, Seyfert subtypes could still form an obscuration sequence, if the dust extinguishing the variable flux component has a grey extinction law \citep{maiolino_dust_2001}. However, obscuration levels in AGN have also been estimated with X-rays. }Most previous studies broadly classified Seyferts into Type 1 and Type 2, and found that most Type 1 Seyferts are X-ray unobscured and most Type 2 Seyferts are \refbftwo{X-ray} obscured \citep{risaliti_absorbing_1999,singh_xray_2011,merloni_bongiorno_2014,davies_torus_2015}. In addition, \citet{shimizu_xraytype_2018} studied a sample selected from a nearly complete sample of AGN up to the Compton thick limit and found that for types 1.0, 1.2, and 1.5, the X-ray absorbed fraction is $<0.12$, $0.08^{+0.08}_{-0.05}$, and $0.05^{+0.16}_{-0.15}$. \refbf{These studies, in conjunction with this work, provide no evidence for grey dust. Both approaches yield similar results, though one caveat is that the extinguishing material in the line-of-sight to the broad line region, to the variable accretion disc, and to the X-ray corona, may not be all the same.}

\section{Discussion and Conclusion}\label{sec:conclusion}
We characterised the optical variability of 246 low-luminosity AGN at $z<0.1$ from 6dFGS using the variability structure function \refbf{on} one-day stacked lightcurves from ATLAS. \refbf{We} used \refbf{recent WiFeS spectra to estimate and subtract} the host galaxy flux. We binned the objects in both luminosity and black hole mass to construct ensemble variability structure functions, then fit power laws and broken power laws to measure the slope, break timescale, and amplitude of the SFs. 
Our main findings are:
\begin{enumerate}
    \item We find slopes less than the DRW expectation of 0.5. The slope increases with BH mass but does not change with luminosity. 
    \item We observe no breaks in the SFs up to 200 days.
    \item Using two-epoch spectra, we show that the SFs mostly continue rising \refbftwo{up to} $\sim 20$ years. 
    \item The variability amplitudes at $\Delta t = 50$\ d anticorrelate with the luminosity and positively correlate with BH mass. The slope dependence on BH mass suggests that the amplitude-mass relation changes with $\Delta t$ and 
    vanishes at $\Delta t \sim 2.6$ days. 
\end{enumerate}
Overall, the observed variability behaviour differs significantly from that reported in studies of high-luminosity AGN. One observation that has only been rarely reported in the literature is the 
slope-BH mass relation. Similar though weaker slope-mass trends have been reported by \citet{simm_pan-starrs1_2016} and \citet{arevalo_universal_2024}, while \citet{caplar_optical_2017} found a dependence of the slope on luminosity and BH mass. To explain this result, \citet{caplar_optical_2017} suggested that quasars could display random walk behaviour at longer timescales and have a steeper behaviour at shorter timescales, and that the timescale of the slope change scales with BH mass. It is important to note that it could very well be the case that the SF does not actually follow a broken power law shape. For example, \citet{arevalo_universal_2024} studied a sample of quasars at rest-frame 2900\AA, and found that the AGN power spectra have shapes corresponding to a bending power law, with high frequency slopes that are steeper than a DRW model (i.e. a high-frequency slope of $\alpha_{\mathrm{H}}=-2$). They found that the bend frequency inversely scales with BH mass with a weaker dependence on the Eddington ratio. Similar to our results, they found a dependence of their slopes on the BH mass, which they attribute to the dependence of the bend timescale on the mass. The BH mass could have affected the part of the power spectrum being probed in the measured time interval, which could have affected the measured slopes. In our SFs, the break timescale is expected to be lower than their sample, so we could be probing the bend of the bending power law without the sufficient baseline to observe the flattening part of the bending power law seen by \citet{arevalo_universal_2024}. Note, that the \citet{arevalo_universal_2024} sample is at a higher luminosity and BH mass, and shorter rest-frame wavelength, so the intrinsic shape of our SFs could be distinct from theirs. Nevertheless, using an inappropriate model to parametrise the SFs may obfuscate the underlying physics behind the variability. 

A slope shallower than 0.5 can alternatively arise from the superposition of multiple stochastic variability processes \citep{kelly_stochastic_2011}. In this framework, the observed trend of the slope with BH mass may suggest that one process becomes progressively dominant at higher masses. If there is one process operating on a characteristic timescale $t_{\mathrm{char}} \sim t_{\mathrm{orb}}$ and another process operating on a distinct timescale, the characteristic timescale inferred from the combined signal would not correspond to that of either individual process. Different variability timescales could also imply multiple break timescales in the SF. The existence of two variability processes has been previously suggested. Recently, \citet{neustadt_using_2022}, pioneered a method that uses multi-wavelength lightcurves to model temperature fluctuations in the accretion disc. In a sample of AGN, they found two signals; fast, outwards travelling waves, and slow, ingoing waves. This result is contrary to the traditional lamppost model, where only the outwards travelling component is expected. This method has been subsequently applied to other AGN \citep{stone_temperature_2023,neustadt_agn_2024}, where some similar results have been found. It is presently unclear how these components map to variability amplitudes on a structure function and how their amplitudes and timescales compare, but it could be an interesting topic for future work. 

If the slope-mass correlation is caused by two variability processes, then, with the absence of a slope-luminosity relation, it may provide important constraints on the nature of the second mechanism. For example, it is generally accepted that the reprocessing of coronal X-rays by the accretion disc contributes to a fraction of the optical variability, though the fraction is unknown. It is possible that this fraction could fluctuate with the size or extent of the corona, which are likely related to the properties of the AGN. For instance, \citet{serafinelli_investigating_2024} suggested that the X-ray corona size increases with BH mass, assuming a typical scale relative to $R_g$. It is also known that the accretion disc and X-ray corona are energetically coupled, as shown by the $L_{\mathrm{X}}-L_{\mathrm{UV}}$ relation \citep{vignali_lxluv_2003,lusso_x-ray_2010}, so the state of the accretion disc will affect that of the corona and vice-versa. According to this relation, at the low-mass, low-luminosity regime we probe, the fraction of X-ray emission relative to the UV continuum is higher, which could increase the fraction of X-rays reprocessed by the accretion disc. In addition, there are studies that suggest the disc structure and accretion processes are affected by the Eddington ratio, such as a transition to a radiatively inefficient flow at extremely low Eddington ratio  \citep{narayan_advection-dominated_1995,hagen_collapse_2024,kang_collapse_2025}. However, if the slope correlation is driven by the Eddington ratio, we would expect to see a trend with the luminosity as well, instead of just with BH mass. 

Setting aside the possibility of multiple variability mechanisms, it is not surprising that the variability behaviour of the cooler outer disc that we probe in the rest-frame optical part of the AGN emission spectrum differs from that of the hotter inner disc well-studied in the literature. Studies have typically found an anticorrelation of the variability amplitude with the rest-frame wavelength \citep[see Table 5 by][and references therein]{de_cicco_structure_2022}, but some studies have found that this anticorrelation flattens at wavelengths longer than $\sim 3000$\AA \ \citep{caplar_optical_2017,tang_universality_2023}, suggesting that the variability behaviour differs between the inner and outer disc. There are also works that suggest that the accretion disc could have multiple components, e.g., \citet{li_twozone_2024} suggested the AGN 1ES 1927+654 consists of a inner overheated slim disc and an outer thin disc. Some instabilities only apply to a certain temperature range and thus a certain region of the disc, such as an instability caused by the iron opacity bump at $\sim 1.8 \times 10^5 \mathrm{K}$ \citep{jiang_opacity-driven_2020}, or instabilities from changes in the disc opacity due to the hydrogen ionisation threshold at 6000 K \citep{cannizzo_accretion_1992,cannizzo_reiff_1992}. Therefore, they may only be observed in a certain part of the disc. 

Even though our results place constraints on the variability properties of low-luminosity AGN, they do not uniquely identify the physical mechanisms driving this variability. While we speculate that the slope-mass correlation could be driven by one variability process dominating over another, and that one process might dominate at high mass and luminosity, our sample does not probe that regime. Previous studies in that parameter space also typically only probe the hotter inner disc. A natural next step is to bridge the gap in parameter space between low- and high-luminosity samples by studying objects at intermediate luminosities in the same rest-frame wavelengths, thereby probing comparable regions of the accretion disc. Upcoming time-domain surveys such as LSST will be particularly powerful in this context, as its multi-band coverage and cadence of a few days will enable detailed studies of how AGN variability depends on wavelength and luminosity across a wide range of physical regimes.
\appendix

\section*{Acknowledgements}
\refbf{The authors thank the reviewer, Andy Lawrence, for helpful comments that improved the quality of the paper.} AHTT and NA were supported by Australian Government Research Training Program (RTP) Scholarship.
This paper is based on observations made with the Australian National University 2.3m Telescope at Siding Springs Observatory. We thank the WiFeS observers Katie Auchettl, Patrick Tisserand and Harrison Abbot for efforts in acquiring some spectra used in this paper. 
The automation of the telescope was made possible through an initial grant provided by the Centre of Gravitational Astrophysics and the Research School of Astronomy and Astrophysics at the Australian National University and through a grant provided by the Australian Research Council through LE230100063. The Lens proposal system is maintained by the AAO Research Data \& Software team as part of the Data Central Science Platform. We acknowledge the traditional custodians of the land on which the telescope stands, the Gamilaraay people, and pay our respects to elders past and present.
This work has made use of data from the Asteroid Terrestrial-impact Last Alert System (ATLAS) project. The ATLAS project is primarily funded to search for near earth asteroids through NASA grants NN12AR55G, 80NSSC18K0284, and 80NSSC18K1575; byproducts of the NEO search include images and catalogs from the survey area. This work was partially funded by Kepler/K2 grant J1944/80NSSC19K0112 and HST GO-15889, and STFC grants ST/T000198/1 and ST/S006109/1. The ATLAS science products have been made possible through the contributions of the University of Hawaii Institute for Astronomy, the Queen’s University Belfast, the Space Telescope Science Institute, the South African Astronomical Observatory, and The Millennium Institute of Astrophysics (MAS), Chile.
The national facility capability for SkyMapper has been funded through ARC LIEF grant LE130100104 from the Australian Research Council, awarded to the University of Sydney, the Australian National University, Swinburne University of Technology, the University of Queensland, the University of Western Australia, the University of Melbourne, Curtin University of Technology, Monash University, and the Australian Astronomical Observatory. SkyMapper is owned and operated by The Australian National University’s Research School of Astronomy and Astrophysics. The survey data were processed and provided by the SkyMapper Team at ANU. The SkyMapper node of the All-Sky Virtual Observatory (ASVO) is hosted at the National Computational Infrastructure (NCI). Development and support of the SkyMapper node of the ASVO has been funded in part by Astronomy Australia Limited (AAL) and the Australian Government through the Commonwealth’s Education Investment Fund (EIF) and National Collaborative Research Infrastructure Strategy (NCRIS), particularly the National eResearch Collaboration Tools and Resources (NeCTAR) and the Australian National Data Service Projects (ANDS).
This research has made use of the NASA/IPAC Infrared Science Archive, which is funded by the National Aeronautics and Space Administration and operated by the California Institute of Technology.

\section*{Data Availability}
The WiFeS spectra will be published in a separate paper. SMSS data underlying this paper are available at the SkyMapper node of the All-Sky Virtual Observatory (ASVO), hosted at the National Computational Infrastructure (NCI) at \href{https://skymapper.anu.edu.au}{https://skymapper.anu.edu.au}. 6dFGS data are available at \href{http://www-wfau.roe.ac.uk/6dFGS/}{http://www-wfau.roe.ac.uk/6dFGS/} and the Final Data Release is available for public access. Data from NASA/ATLAS are publicly available at \href{https://fallingstar-data.com/forcedphot/}{https://fallingstar-data.com/forcedphot/}




\bibliographystyle{mnras}
\bibliography{reference} 




\appendix
\begin{figure*}
    \centering
    \includegraphics[width=\linewidth]{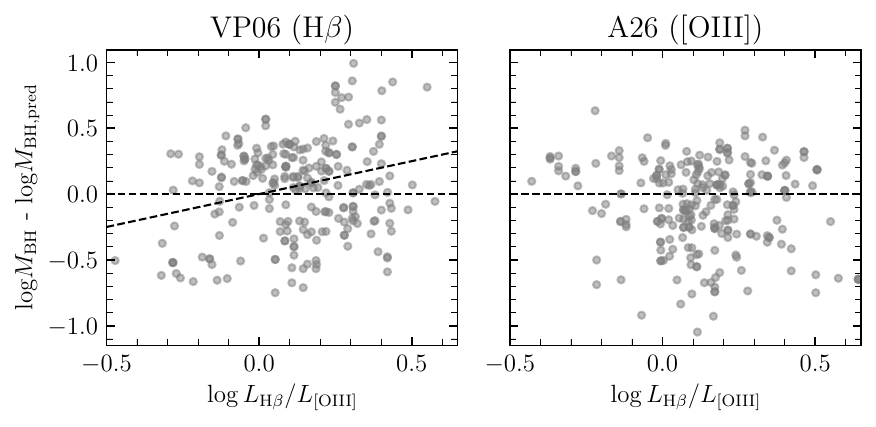}
    \renewcommand{\thefigure}{A1}
    \caption{Mass residuals of the slope-mass correlation vs. $\log L_{\mathrm{H\beta}}/L_{\mathrm{[OIII]}}$. In the left panel, the \citet{vestergaard_peterson_2006} H$\mathrm{\beta}$ mass calibration is used, while the [OIII] mass calibration by \citet{amrutha_masscorrection_2026} is used in the right panel. A dashed line with a slope of 0.5 is plotted on the left panel.}
    \label{fig:mass_residual}
\end{figure*}

\section{Biases in single-epoch masses} \label{appendix:masses}

\citet{amrutha_masscorrection_2026} recently found that single-epoch mass estimates are biased by the \ratio \ ratio. 
The majority of BH mass estimates are derived using the single-epoch virial mass method, which relies on the radius-luminosity (R-L) relation calibrated using a small sample of low-redshift AGN through reverberation mapping \citep[e.g.][]{bentz_rl_2013}. However, BH masses estimated through this method have an uncertainty of $\sim0.5$ dex \citep{shen_mass_2013}. One source of uncertainty is the geometry of the BLR, which is represented by a dimensionless virial factor that is assumed to be constant for all AGN in the single-epoch method. Moreover, this method assumes that the BLR exhibits virial breathing behaviour, which means that the responsivity weighted BLR radius increases in response to an increase in the ionising continuum and vice versa, but some studies have found no breathing behaviour in individual AGN or samples of AGN \citep{wang_rm_2020,jiang_breathing_2025}. In addition, secondary correlations of the R-L relation with parameters such as the Eddington ratio and \ratio\ have been found. One recent work, \citet{amrutha_masscorrection_2026}, found that a complete sample of AGN at $z < 0.1$ varied in luminosity and mass estimates by a factor of two on average after twenty years, while the line widths remained constant on average. Since the narrow line region is located at kpc scales, short term variability is expected to not affect $L_{\mathrm{[OIII]}}$, so this implies that objects with high \ratio\ have overestimated BH masses and vice versa. Therefore, they suggested that using the narrow [OIII] line luminosity instead of the broad H$\beta$ line luminosity yields more accurate BH masses. They provide the following equation for the [OIII] derived black hole mass derived using the [OIII] R-L relation by \citet{wang_woo_rl_2024}, 
\begin{equation}
    \log M_{\mathrm{BH}} = 2 \log (\mathrm{\frac{FWHM_{H\beta}}{1000 kms^{-1}}}) + \beta \log(\frac{L_{\mathrm{[OIII]}}}{10^{41.69} \mathrm{ergs^{-1}}}) + C,
\end{equation}
where $\beta = 0.49\pm0.04$ and $C = 6.85\pm0.17$. 

Assuming that the slope-mass dependence doesn't change with the AGN activity level probed by \ratio, we can test whether the [OIII] masses are more accurate than the H$\beta$ masses. We use the [OIII] mass and $L_{5100}$ to bin the SFs, then repeat the process detailed in Section \ref{subsec:slope_method} to measure the slope-BH mass dependence. Then, we compare the residuals from the best fit slope-mass correlation, assuming the the SF slope only depends on the BH mass. Objects with high \ratio\ have overestimated BH masses, so they are expected to lie below the slope-mass relation and vice versa. With the best fit relation, we can calculate a predicted mass given the measured SF slope, $\log M_{\mathrm{BH, pred}}$, and define a mass residual as $\log M_{\mathrm{BH}}-\log M_{\mathrm{BH, pred}}$. We can also predict the behaviour of the mass residuals with \ratio. The slope of the R-L relation is $\sim 0.5$, so if the deviation of the slope from the slope-mass relation is mainly due to a secondary correlation of the R-L relation with \ratio, the slope of the mass residuals plotted against $\log L_{\mathrm{H\beta}}/L_{\mathrm{[OIII]}}$ should approximately follow this slope. Conversely, the [OIII] mass is suggested to be unaffected by \ratio, so the mass residuals calculated using the [OIII] mass should have a slope of zero. Figure \ref{fig:mass_residual} shows the mass residuals of the H$\mathrm{\beta}$ and [OIII] mass. In the left panel, the points approximately follow the expected slope of 0.5, suggesting that the mass residuals are biased by \ratio, while the points on the right panel have an approximately flat distribution, suggesting that [OIII] mass estimates are not biased by the short-term AGN variability and yields more accurate BH masses. Note, that the ensemble SFs are binned in bins of mass and luminosity but not \ratio\ due to the limited sample size, which introduces scatter into the points. 


\bsp	
\label{lastpage}
\end{document}